\documentclass[aps,prb,preprint,groupedaddress]{revtex4}
\usepackage{graphicx}
\begin{document}

\title{Dynamics of Entanglement for One-Dimensional Spin Systems in an 
External Time-Dependent Magnetic Field}

\author{Zhen Huang and 
Sabre Kais\footnote{
Corresponding author:  kais@purdue.edu\\Accepted for publication in the Int. J. Quantum Information, Sept. 2005}}

\affiliation{Department of Chemistry, Purdue University, West Lafayette, IN 47907}

\begin{abstract}
\noindent

We study the dynamics  of entanglement  
for the XY-model, one-dimensional spin systems coupled 
through nearest neighbor exchange interaction 
and subject to  an external time-dependent magnetic field. Using the two-site 
density matrix, we calculate the time-dependent entanglement of formation between 
nearest neighbor qubits. We investigate the effect of varying the temperature,
the anisotropy parameter and the external 
time-dependent magnetic field on the entanglement.  
We have found that the entanglement can be localized between 
nearest neighbor qubits for certain values of the 
external time-dependent magnetic field. 
Moreover, as known for the magnetization of this model, 
the entanglement shows  nonergodic behavior, it does not approach its
equilibrium value at the infinite time limit.

\end{abstract}

\maketitle
\clearpage

\section{introduction}

Quantum entanglement is regarded as the resource of quantum 
information processing with no classical analog\cite{diviccezo,entg1,Nielsen,gruska}.
The corresponding investigation is  currently a very active area of research
\cite{vedral,hill,Wootters98,vidal,ebr,blatt} due to its potential
applications in quantum communication, such as quantum teleportation 
\cite{entg2,dik}, superdense coding \cite{entg3}, quantum key distribution 
\cite{entg4}, telecoloning \cite{entg5} and decoherence  
in quantum computers\cite{div,whaley}.\\
 
Multiparticles systems are the central interest in the 
field of quantum information, in particular, the quantification of
 the entanglement contained in quantum states, because the entanglement 
 is the physical resource to perform some of the most important quantum 
 information tasks, like quantum information transfer or quantum computation. 
 Osterloh et.al\cite{Osterloh02} connected the theory of critical phenomena 
 with quantum information by exploring the entangling resources of a system 
 close to quantum critical point in a class of one-dimensional magnetic systems.

Recently\cite{omar}, we have demonstrated that for a class of 
one-dimensional magnetic systems entanglement can be controlled and
tuned by varying the anisotropy parameter in the XY Hamiltonian 
and by introducing impurities into the systems in the equilibrium state. 
However, offering a potentially ideal protection against 
environmentally induced decoherence is difficult in information 
encoding and readout. An important motivation  to study the 
dynamics of entanglement while varying an external magnetic field is 
to investigate whether it is possible to protect the entanglement 
from the effects of environment such as an external magnetic field and
change of temperature. \\

Amico et. al\cite{Amico04} study the dynamics of entanglement in one-dimensioanl spin systems 
using Ising-type models. They analyze the time evolution of initial Bell states created in a
fully polarized background and on the ground state. They have found that the 
pairwise entanglement propagates with a velocity proportional to the reduced interaction for 
all the four Bell states. Moreover, they show that the 
"entanglement  wave" evolving from a Bell state on the ground state turns out to be
very localized in space time.

In this paper, we consider the dynamics of a set of localized spin-1/2 particles 
coupled through an exchange interaction and subject to an external time-dependent 
magnetic field. In Sec. II, we introduce the Liouville equation 
for the density matrix and present the numerical solution of 
the general XY-model in a lattice with $N$ sites in an external 
 time-dependent magnetic field $h(t)$. In Sec. III, the solutions presented 
 in the previous section are used to compute the magnetization and 
 spin-spin correlation functions. The entanglement of formation is 
 briefly introduced in Sec. IV and expressed in terms of the different
 spin-spin correlation functions. Finally, Sec. V is devoted to the results
 and discussions.

\section{Solution of the time-dependent XY model}

In this section, we present the numerical solution of the XY model for a 
one-dimensional lattice with $N$ sites in an external 
 time-dependent magnetic field $h(t)$. 
The Hamiltonian for such a chain of interacting spins, with nearest-neighbor 
interaction only,  is given by

\begin{equation}\label{initialHamiltonian}
H = -\frac{J}{2}(1+\gamma)\sum_{i=1}^N\sigma_i^x \sigma_{i+1}^x -
\frac{J}{2}(1-\gamma)\sum_{i=1}^N \sigma_i^y \sigma_{i+1}^y -
\sum_{i=1}^N h(t) \sigma_i^z \, ,
\end{equation}

\noindent where $J$ is the coupling constant, $\sigma^a$ are the Pauli matrices 
($a = x,y,z$), and $\gamma$ is the degree of anisotropy. 
We set $J=1$ for convenience. The periodic boundary condition is cyclic, namely,   
$\sigma_{N+1}^a=\sigma_1^a$.

The standard procedure used to solve Eq.(\ref{initialHamiltonian}) 
is to transform the spin operators into
fermionic operators\cite{Lieb,Barouch70}. Let us define  the raising and lowering 
operators $a_i^+$, $a_i$,
\begin{equation}
a_i^+=\frac{1}{2}(\sigma_i^x + i\sigma_i^y),~~~~
a_i=\frac{1}{2}(\sigma_i^x - i\sigma_i^y),
\end{equation}
\noindent in terms of which the Pauli matrices are given by
\begin{equation}
\sigma_i^x=a_i^+ + a_i,~~~~\sigma_i^y=\frac{a_i^+ - a_i}{i},~~~~\sigma_i^z=2a_i^+ a_i - I.
\end{equation}
\noindent These operators can be expressed in terms of Fermi operators $b_i$, $b_i^+$ 
\begin{equation}
a_i=exp(-\pi i\sum_{j=1}^{i-1} b_j^+ b_j)b_i,~~~~
a_i^+=b_i^+ exp(\pi i\sum_{j=1}^{i-1} b_j^+ b_j).
\end{equation}
\indent Next, we introduce the Fourier transform for a general $h(t)$
\begin{equation}
b_{j}^{+}=\frac{1}{\sqrt{N}}\sum_{p=-N/2}^{N/2}exp(ij\phi_p)c_p^+~,~~~~
b_{j}=\frac{1}{\sqrt{N}}\sum_{p=-N/2}^{N/2}exp(-ij\phi_p)c_p~,
\end{equation}
\noindent where $\phi_p=\frac{2\pi p}{N}$. Thus, the Hamiltonian assumes the following form
\begin{equation}\label{finalHamiltonian}
H=\sum_{p=1}^{N/2}{\alpha_p(t)[c_p^{+}c_p+c_{-p}^{+}c_{-p}]+i\delta_p[c_p^{+}c_{-p}^{+}+c_{p}c_{-p}]+2h(t)}~.
\end{equation}

\noindent where $~\alpha_p(t)=-2cos\phi_p-2h(t)~$ and 
$~\delta_p=2\gamma sin\phi_p~$. Since $~[\tilde{H}_p,\tilde{H}_q]=0~$,
 we can write Eq. (\ref{finalHamiltonian}) as
\begin{equation}
H=\sum_{p=1}^{N/2}\tilde{H}_p,
\end{equation}
\noindent with 
\begin{equation}
\tilde{H}_p=\alpha_p(t)[c_p^{+}c_p+c_{-p}^{+}c_{-p}]+
i\delta_p[c_p^{+}c_{-p}^{+}+c_{p}c_{-p}]+2h(t)~.
\end{equation}
This means that the space of $\tilde{H}$ can be decomposed into noninteracting 
subspaces, each of four dimensions. 
Using the following basis for the $\emph{p}$th subspace:
\begin{equation}\label{basisSet}
(|0>; c_p^+c_{-p}^+|0>; c_p^+|0>; c_{-p}^+|0>) ,
\end{equation}
\noindent we can explicitly obtain
\begin{equation}\label{subspaceHamiltonian}
\tilde{H}_p(t)=\scriptsize{
\left(\begin{array}{cccc}
2h(t)&-i\delta_p&0&0\\
i\delta_p&-4cos\phi_p-2h(t)&0&0\\
0&0&-2cos\phi_p&0~\\
0&0&0&-2cos\phi_p
\end{array}\right)}.
\end{equation}
\indent In this paper, the initial condition chosen at 
$t=0$ is  thermal equilibrium of the system, namely, 
the density matrix of 
the $\emph{p}$th subspace at time $t$ $\rho_p(t)$ is given by
\begin{equation}
\rho_p(0)=e^{-{\beta}\tilde{H}_p(0)}, \;\;\;
\beta=1/kT
\end{equation}
\indent $k$ is the Boltzmann constant. 
Therefor, using Eq. (\ref{subspaceHamiltonian}), we obtain
\begin{equation}\label{rho0}
\rho_p(0)= e^{2{\beta}cos\phi+2{\beta}\Lambda[h(0)]} \scriptsize{
\left(\begin{array}{cccc}
k_{11}^p&k_{12}^p&0&0\\
k_{21}^p&k_{22}^p&0&0\\
0&0&k_{33}^p&0~\\
0&0&0&k_{44}^p
\end{array}\right)}.
\end{equation}
where
\begin{equation}
\Lambda[h(0)]=\{[cos\phi+h(0)]^2+\gamma^{2}sin^2\phi\}^{1/2}~,
\end{equation}
and the matrix elements are given by
\begin{equation}
k_{11}^p=\frac{\{\Lambda[h(0)]+cos\phi+h(0)\}e^{-4\beta\Lambda[h(0)]}+\{\Lambda[h(0)]-cos\phi-h(0)\}}{2\Lambda[h(0)]}~,
\end{equation}
\begin{equation}
k_{12}^p=\frac{i\delta\{1-e^{-4\beta\Lambda[h(0)]}\}}{4\Lambda[h(0)]}~,
\end{equation}
\begin{equation}
k_{21}^p=\frac{-i\delta\{1-e^{-4\beta\Lambda[h(0)]}\}}{4\Lambda[h(0)]}~,
\end{equation}
\begin{equation}
k_{22}^p=\frac{\{\Lambda[h(0)]-cos\phi-h(0)\}e^{-4\beta\Lambda[h(0)]}+\{\Lambda[h(0)]+cos\phi+h(0)\}}{2\Lambda[h(0)]}~,
\end{equation}
\begin{equation}
k_{33}^p=k_{44}^p=e^{-2 \beta \Lambda[h(0)]}~.
\end{equation}
Let $U_p(t)$ be the time-evolution matrix in the $\emph{p}$th subspace, then
\begin{equation}\label{timeEvolutionMatrix}
i\frac{dU_p(t)}{dt}=U_p(t)\tilde{H}_p(t)~, \;\;\; \hbar=1
\end{equation}
Since $\tilde{H}_p(t)$ is in a block diagonal form
\begin{equation}
U_p(t)=\scriptsize{
\left(\begin{array}{cccc}
U_{11}^p&U_{12}^p&0&0\\
U_{21}^p&U_{22}^p&0&0\\
0&0&U_{33}^p&0~\\
0&0&0&U_{44}^p
\end{array}\right)}~,
\end{equation}
\noindent where the upper-left block is determined from
\begin{equation}
i\frac{d}{dt}\scriptsize{\left(\begin{array}{cc}
U_{11}^p&U_{12}^p\\U_{21}^p&U_{22}^p\end{array}\right)}=
\scriptsize{\left(\begin{array}{cc}U_{11}^p&U_{12}^p\\U_{21}^p&U_{22}^p\end{array}\right)}
\scriptsize{\left(\begin{array}{cc}2h(t)&-i\delta_p\\i\delta_p
&-4cos\phi_p-2h(t)\end{array}\right)}.
\end{equation}

\indent Thus, the Liouville equation of the system which is given by
\begin{equation}
i\frac{d\rho(t)}{dt}=[H(t),\rho(t)]
\end{equation}
 can be solved exactly because it can be  decomposed into uncorrelated subspaces.
 In the $\emph{p}th$ subspace, the solution of Liouville equation is
\begin{equation}\label{liouvilleEquation}
\rho_p(t)=U_p(t)\rho_p(0)U_p(t)^\dagger~.
\end{equation}  
\noindent In this study the magnetic field will be presented by a
step function of the form,
\begin{displaymath}
h(t)=\left\{ 
\begin{array}{ll}
a &~~~~ t\le0\\
b &~~~~ t > 0
\end{array}
\right\}
\end{displaymath}

\noindent which will allow us to obtain  the 
solution of Eq. (\ref{timeEvolutionMatrix}) 
\begin{equation}
U_p(t)= e^{2itcos\phi} \scriptsize{
\left(\begin{array}{cccc}
U_{11}^p&U_{12}^p&0&0\\
U_{21}^p&U_{22}^p&0&0\\
0&0&U_{33}^p&0~\\
0&0&0&U_{44}^p
\end{array}\right)}~,
\end{equation}
\noindent where 

\begin{equation}
U_{11}^p=\frac{-i(cos\phi+b)sin[2t\Lambda(b)]}{\Lambda(b)}+cos[2t\Lambda(b)]~,
\end{equation}
\begin{equation}
U_{12}^p=\frac{-\delta sin[2t\Lambda(b)]}{2\Lambda(b)}~,
\end{equation}
\begin{equation}
U_{21}^p=\frac{\delta sin[2t\Lambda(b)]}{2\Lambda(b)}~,
\end{equation}
\begin{equation}
U_{22}^p=\frac{i(cos\phi+b)sin[2t\Lambda(b)]}{\Lambda(b)}+cos[2t\Lambda(b)]~,
\end{equation}
\begin{equation}
U_{33}^p=U_{44}^p=1~.
\end{equation}

\noindent From Eq. (\ref {liouvilleEquation})  we can get 
\begin{equation}\label{rhot}
\rho_p(t)=e^{2{\beta}cos\phi+2{\beta}\Lambda[h(0)]} \scriptsize{
\left(\begin{array}{cccc}
\rho_{11}^p&\rho_{12}^p&0&0\\
\rho_{21}^p&\rho_{22}^p&0&0\\
0&0&\rho_{33}^p&0~\\
0&0&0&\rho_{44}^p
\end{array}\right)}~,
\end{equation}
\noindent where the matrix elements are given by

\begin{equation}
\rho_{11}^p=\frac{\{\delta^2(b-a)sin^2[2t\Lambda(b)]+\zeta\}e^{-4\beta\Lambda(a)}+\delta^2(a-b)sin^2[2t\Lambda(b)]+\eta}{4\Lambda^2(b)\Lambda(a)}
\end{equation}
\begin{equation}
\rho_{12}^p=\frac{\delta(1-e^{-4\beta\Lambda(a)})\{\Lambda(b)sin[4t\Lambda(b)](b-a)+i\{\Lambda^2(b)+2(a-b)(cos\phi+b)sin^2[2t\Lambda(b)]\}\}}{4\Lambda^2(b)\Lambda(a)}
\end{equation}
\begin{equation}
\rho_{22}^p=\frac{\{\delta^2(a-b)sin^2[2t\Lambda(b)]+\eta\}e^{-4\beta\Lambda(a)}+\delta^2(b-a)sin^2[2t\Lambda(b)]+\zeta}{4\Lambda^2(b)\Lambda(a)}
\end{equation}
\begin{equation}
\rho_{21}^p=(\rho_{12}^p)^*
\end{equation}
\begin{equation}
\rho_{33}^p=\rho_{44}^p=e^{-2\beta\Lambda(a)}
\end{equation}

with 

\begin{equation}
\zeta=2\Lambda^2(b)(\Lambda(a)+cos\phi+a)
\end{equation}
\begin{equation}
\eta=2\Lambda^2(b)(\Lambda(a)-cos\phi-a)
\end{equation}

\section{MAGNETIZATION AND SPIN-SPIN CORRELATION FUNCTIONS}
\indent The magnetization in the XY model is defined as 
\begin{equation}\label{magnetizationDefinition}
M=\frac{1}{N}\sum_{j=1}^N S_j^z~,
\end{equation}
\noindent which can be written in terms of the operators~$c_p^+$~, $c_p^-$~as
\begin{equation}
M=\frac{1}{N}\sum_{p=1}^{N/2}M_p,
\end{equation}
\noindent where $M_p=c_p^+c_p+c_{-p}^+c_{-p}-1$~. So we can get the z-direction magnetization
\begin{equation}\label {magnetization}
M_z(t)=\frac{1}{N}\frac{Tr[M\rho]}{Tr[\rho]}=\frac{1}{N}\sum_{p=1}^{N/2}\frac{Tr[M_p\rho_p(t)]}{Tr[\rho_p(0)]}.
\end{equation}
\noindent Using Eqs. (\ref{basisSet}),(\ref{rho0}) and (\ref{rhot})
Eq. (\ref{magnetization}) gives
\begin{equation}
M_z(t)=\frac{1}{4N}\sum_{p=1}^{N/2}\frac{tanh[\beta\Lambda(a)]\{2\delta^2(b-a)sin^2[2t\Lambda(b)]+4\Lambda^2(b)(cos\phi+a)\}}{\Lambda^2(b)\Lambda(a)}~.
\end{equation}
\indent The three instantaneous spin-spin correlation functions are defined as
\begin{equation}
S_{lm}^x=<S_l^xS_m^x>~,\;\;
S_{lm}^y=<S_l^yS_m^y>~,\;\;
S_{lm}^z=<S_l^zS_m^z>
\end{equation}

\noindent Lieb Schultz and Mattis (LSM) \cite{Lieb} show that 
\begin{equation}
\label{spinCorrelaionx}
S_{lm}^x=\frac{1}{4}<B_lA_{l+1}B_l...A_{m-1}B_{m-1}A_m>~,
\end{equation}
\begin{equation}
\label{spinCorrelationy}
S_{lm}^y=\frac{1}{4}(-1)^{l-m}<A_lB_{l+1}A_{l+1}B_{l+2}...B_{m-1}A_{m-1}B_m>~,
\end{equation}
\begin{equation}
\label{spinCorrelationz}
S_{lm}^z=\frac{1}{4}<A_lB_lA_mB_m>~,
\end{equation}
\noindent where 
\begin{equation}
A_i=b_i^++b_i \;;\;\;\;
B_i=b_i^+-b_i.
\end{equation}
These three correlation functions are given as expectation 
values of products of fermion operators. Using the Wick \cite{Wick50} 
theorem, the expressions can be expressed as Pfaffians $(pf)$. 
In particular, we have
\begin{equation}
S_{lm}^x=\frac{1}{4}pf\scriptsize{
\left(\begin{array}{ccccccc}
<B_lA_{l+1}>&<B_lB_{l+1}>&\cdot&\cdot&\cdot&<B_lB_{m-1}>&<B_lA_m>\\
~&<A_{l+1}B_{l+1}>&\cdot&\cdot&\cdot&<A_{l+1}B_{m-1}>&<A_{l+1}A_m>\\
~&~&\cdot&\cdot&\cdot&\cdot&\cdot\\
~&~&~&\cdot&\cdot&\cdot&\cdot\\
~&~&~&~&\cdot&\cdot&\cdot\\
~&~&~&~&~&<A_{m-1}B_{m-1}>&<A_{m-1}A_m>\\
~&~&~&~&~&~&<B_{m-1}A_m>
\end{array}\right)}
\end{equation}

\begin{equation}
S_{lm}^y=\frac{(-1)^{l-m}}{4}pf\scriptsize{
\left(\begin{array}{ccccccc}
<A_lB_{l+1}>&<A_lA_{l+1}>&\cdot&\cdot&\cdot&<A_lA_{m-1}>&<A_lB_m>\\
~&<B_{l+1}A_{l+1}>&\cdot&\cdot&\cdot&<B_{l+1}A_{m-1}>&<B_{l+1}B_m>\\
~&~&\cdot&\cdot&\cdot&\cdot&\cdot\\
~&~&~&\cdot&\cdot&\cdot&\cdot\\
~&~&~&~&\cdot&\cdot&\cdot\\
~&~&~&~&~&<B_{m-1}A_{m-1}>&<B_{m-1}B_m>\\
~&~&~&~&~&~&<A_{m-1}B_m>
\end{array}\right)}
\end{equation}

\begin{equation}
S_{lm}^z=\frac{1}{4}pf\scriptsize{
\left(\begin{array}{ccc}
<A_lB_l>&<A_lA_m>&<A_lB_m>\\
~&<B_lA_m>&<B_lB_m>\\
~&~&<A_mB_m>
\end{array}\right)}
\end{equation}
\noindent where 

{\setlength\arraycolsep{2pt}
\begin{eqnarray}
<B_lA_m>&=&\frac{1}{N}\sum_{p=1}^{N/2}\frac{1}{\Lambda^2(b)
\Lambda(a)[1+e^{-2\beta\Lambda(a)}]^2}\{sin(\frac{2{\pi}(m-l)p}{N})\{\Lambda^2(b){}
\nonumber\\
&&{}+2(a-b)(cos\phi+b)sin^2[2t\Lambda(b)]\}+cos(\frac{2{\pi}(m-l)p}{N}){}
\nonumber\\
&&{}\{\delta^2(b-a)sin^2[2t\Lambda(b)]+2\Lambda^2(b)(cos\phi+a)\}[1-e^{-4\beta\Lambda(a)}]\}~,
\end{eqnarray}
}
\begin{equation}
<A_lA_m>=\frac{1}{N}\sum_{p=1}^{N/2}\{2cos(\frac{2\pi (m-l)p}{N})+
\frac{i\delta(a-b)sin(\frac{2\pi (m-l)p}{N})sin[4t\Lambda(b)]tanh(\beta\Lambda(a))}{\Lambda(b)\Lambda(a)}\}~, 
\end{equation}
\begin{equation}
<B_lB_m>=\frac{1}{N}\sum_{p=1}^{N/2}\{-2cos(\frac{2\pi (m-l)p}{N})+\frac{i\delta(a-b)sin(\frac{2\pi(m-l) p}{N})sin[4t\Lambda(b)]tanh(\beta\Lambda(a))}{\Lambda(b)\Lambda(a)}\}~. 
\end{equation}

\section{ENTANGLEMENT OF FORMATION}

\indent The concept of entanglement of formation is related to the amount of entanglement needed to prepare the state $\rho$, where~$\rho$~is the density matrix. It was shown by Wootters\cite{Wootters98} that
\begin{equation}
E(\rho)=\mathcal{E}(C(\rho)),
\end{equation}
\noindent where the function $\mathcal{E}$ is given by
\begin{equation}
\mathcal{E}=h(\frac{1+\sqrt{1-C^2}}{2}),
\end{equation}
where $h(x)=-xlog_2x-(1-x)log_2(1-x)$ and the 
concurrence $\textit{C}$ is defined as
\begin{equation}
C(\rho)=max\{0,~\lambda_1-\lambda_2-\lambda_3-\lambda_4~\}.
\end{equation}
 For a general state of two qubits, 
 $\lambda_i$'s are the eigenvalues, in decreasing order, of the 
 Hermitian matrix
\begin{equation}
R \equiv~\sqrt{\sqrt{\rho}~\tilde{\rho}~\sqrt{\rho}},
\end{equation}
\noindent where~$\rho$~is the density matrix and ~$\tilde{\rho}$~ is 
the spin-flipped state defined as\\
\begin{equation}
\tilde{\rho}=(\sigma_y \otimes \sigma_y) \rho^*(\sigma_y \otimes \sigma_y).
\end{equation}

Alternatively, the $\lambda_i$'s are the square roots of the eigenvalues 
of the non-Hermitian $\rho \tilde{\rho}$. Since the density matrix $\rho$ 
follows from the symmetry properties of the Hamiltonian, the $\rho$ must be real and symmetrical\cite{Osterloh02}, plus
the global phase flip symmetry of Hamiltonian, which implies that
$[\sigma^z_i\sigma^z_j,\rho] = 0$,
we obtain 
\begin{equation}
\rho={\left(\begin {array}{cccc} \rho_{1,1} 
& 0 & 0 & \rho_{1,4} \\ 
0 & \rho_{2,2} & \rho_{2,3} & 0 \\
 0 & \rho_{2,3} & \rho_{3,3} & 0\\ 
\rho_{1,4} & 0 & 0 & \rho_{4,4}\end {array} \right)},\\
\end{equation}

\noindent with

\begin{equation}\label{eigenvalue}
\lambda_a = \sqrt{\rho_{1,1} \rho_{4,4}} + | \rho_{1,4}|,~ 
\lambda_b= \sqrt{\rho_{2,2} \rho_{3,3}} + | \rho_{2,3} |,~
\lambda_c = | \sqrt{ \rho_{1,1} \rho_{4,4} } - | \rho_{1,4} | |,~ 
\lambda_d = | \sqrt{\rho_{2,2} \rho_{3,3}} - | \rho_{2,3} |.
\end{equation}

Using the definition $<A>=Tr(\rho A)$, we can express all the 
matrix elements in the density matrix 
in terms of the different spin-spin correlation functions:

\begin{center}
\begin{equation}
\rho_{1,1}=\frac{1}{2} M_l^z + \frac{1}{2} M_m^z + S_{lm}^z + \frac{1}{4},
\end{equation}
\begin{equation}
\rho_{2,2}=\frac{1}{2} M_l^z -\frac{1}{2} M_m^z - S_{lm}^z + \frac{1}{4},
\end{equation}
\begin{equation}
\rho_{3,3}=\frac{1}{2} M_m^z -\frac{1}{2} M_l^z - S_{lm}^z + \frac{1}{4},
\end{equation}
\begin{equation}
\rho_{4,4}= - \frac{1}{2} M_l^z -\frac{1}{2} M_m^z + S_{lm}^z + \frac{1}{4},
\end{equation}
\begin{equation}
\rho_{2,3}= S_{lm}^x + S_{lm}^y,
\end{equation}
\begin{equation}
\rho_{1,4}= S_{lm}^x - S_{lm}^y.
\end{equation}
\end{center}

\section{RESULTS AND DISCUSSIONS}

Our goal is to examine the dynamics  of entanglement in the presence 
of varying  external magnetic field, temperature and 
the anisotropy parameter $\gamma$. First we describe the dynamics for
the Ising model with $\gamma=1$. For a constant magnetic field, it is convenient to
define a dimensionless coupling constant $\lambda=J/h$. This model is known to 
undergo a quantum phase transition at $\lambda_c=1$. The magnetization 
$<\sigma^x>$ is different from zero for $\lambda > 1$ and it vanishes at the 
transition\cite{tobias}. However, the  magnetization  $<\sigma^z>$  is different from zero
for any value of $\lambda$. At the quantum phase transition the correlation length
diverges as $\xi \sim |\lambda-\lambda_c|^{-1}$. When $\lambda \rightarrow 0$, the ground
state becomes a product of spins pointing in the positive $z$-direction. However, 
in the limit $\lambda \rightarrow \infty$, the ground state becomes again 
a product of spins pointing in the positive $x$-direction. In both limits the ground
state approaches a product state, thus the entanglement vanishes. When   $\lambda=1$, 
a fundamental transition in the form of the ground state occurs and the system develops
a nonzero magnetization $<\sigma^x> \neq 0$ which grows as $\lambda$ is increased. The 
calculations of entanglement show that it rises from zero in the two limits
 $\lambda \rightarrow 0$ and $\lambda \rightarrow \infty$ to a maximum value near  the 
 critical point $\lambda_c=1$. Moreover, the range of entanglement, that is the maximum distance between 
 two spins at which the concurrence is different from zero, vanishes unless the two sites are at most 
 next-nearest neighbors.

In Fig (1) we show how the nearest
neighbor concurrence $C(i,i+1)$ evolves with time when the initial $C(i,i+1)$ close to the 
maximum. We choose the parameters $a=b=1.001$, $a=b=0.5$ and the step function
with $a=1.001$ and $b=0.5$. Thus at  $t=0$,  $\lambda$ close to one and  $C(i,i+1)$ close
to maximum. As time evolve, $C(i,i+1)$ oscillate, but it does not reach it is equilibrium 
value at $t \rightarrow \infty$. Barouch et. al. \cite{Barouch70} 
have shown the nonergodic behavior of the 
 the magnetization for
the XY-model. The limit $t \rightarrow \infty$  of the magnetization does not approach its
equilibrium value. This phenomenon, the magnetization is not an ergodic observable in this model
was discusses earlier by Mazur\cite{mazur}. The concurrence $C(i,i+1)$ shows a similar behavior, 
that is nonergodic, since it is related to the 
magnetization and spin-spin correlation functions.  
In the lower panel of Fig. (1), we calculate the thermal
nearest neighbor concurrence $C(i,i+1)$ as a function of time $t$ for $kT=0.5$ and $kT=1.0$. 
For this model, the entanglement is nonzero only in a certain region in the $(kt-\lambda)$
plane\cite{tobias}. The entanglement is largest in the vicinity of the critical point $\lambda_c=1$ and
$kT=0$, this is the quantum critical regime.
As expected the concurrence decreases with increasing temperature at $\lambda$ close to one 
and the oscillations disappeared
at $kT=1.0$.

In Fig.(2) we show the  nearest
neighbor concurrence $C(i,i+1)$ as a function of time $t$ at $kT=0$ and $kT=0.5$ for 
different parameters of the magnetic field, a step function with an initial field
$a=0.5$ and a final field $b=5.0$. For $a=b=0.5$, $\lambda=2 > \lambda_c=1$ and
for  $a=b=5.0$, $\lambda=0.2 < \lambda_c$.
One can see that the concurrence 
starts oscillations when the external 
magnetic field is applied and reaches a limit  
when $~t\to\infty~$, which is again not the equilibrium limit. \\

In order to investigate the property of concurrence
at equilibrium, we calculate the three spin-spin correlation functions as defined in 
Eq.(\ref{spinCorrelaionx}), Eq.(\ref{spinCorrelationy}), 
Eq.(\ref{spinCorrelationz}) and the magnetization in
Eq. (\ref{magnetizationDefinition}). 
Figures (3) and (4)  show the behavior of three spin-spin correlation functions 
and the magnetization as a function of time $t$ at $kT=0$ and $kT=0.5$ respectively.
As reported  by Barouch et. al.\cite{Barouch70}, the magnetization of the Ising model does not 
approach  the equilibrium state limit. Furthermore, 
we find that the three spin-spin correlation functions do not approach 
the  equilibrium state at $t\to\infty$.

To show the effect of the initial and final external magnetic
 field strengths on the entanglement with $t\to\infty$, we show in Fig. (5) the nearest neighbor 
 concurrence $C(i,i+1)$ as a function of the parameters $a$ and $b$ at  $kT=0$ and $\gamma=1$. 
 For $a < 1$ region, the concurrence increases very fast near $b = 1$ and reaches 
a  limit $C(i,i+1) \sim 0.125$  when $~b\to\infty~$. It is surprising that
 the concurrence will not 
 disappear when $b$ increases with $a < 1$. This indicates that the 
 concurrence will not disappear as the final external magnetic field increase 
 at infinite time. It shows that this model is not in agreement with the 
 obvious physical intuition, since we expect that increasing the external 
 magnetic field will destroy the spin-spin correlations functions and make 
 the concurrence vanishes. In our previous calculations\cite{Huang04}, 
 we have found that the concurrence approached a maximum 
 when the external magnetic 
 field is close to the critical point. The concurrence approaches 
 maximum $C(i,i+1) \sim 0.258$
 at $(a=1.37,b=1.37)$,  and decreases 
 rapidly as $a \neq b$. This indicates that the fluctuation of the external 
 magnetic field near the equilibrium state will rapidly  destroy the entanglement. 
 However,in the region where $a > 2.0$, 
 the concurrence is close to zero when  $ b < 1.0$ and maximum close to $1$. Moreover, 
it disappears in the limit of $b\to\infty$.\\
 
 Recently, it was reported that the nearest neighbor concurrence will decrease as 
the temperature increases \cite{Amico04} at the equilibrium state. 
It is interesting to investigate the effect of temperature on the concurrence 
in our model. Fig. (6) shows the nearest neighbor concurrence $C(i,i+1)$ as the parameters
$a$ and $b$ varies at  $kT = 1$. The concurrence in the region where $a< 1$ disappears, 
and the sharp peak shown at $kT=0$ decrease by increasing the temperature. 
The maximum $C(i,i+1) \sim 0.195 $  occurs at $(a=1.76,b=3.0)$.

We now move to consider the dynamics for the anisotropy parameter $\gamma \neq 1$.
In Fig.(7) we show the  nearest
neighbor concurrence $C(i,i+1)$ as a function of time $t$ at $kT=0$ for 
the same parameters of the magnetic field  shown in Fig. (2) for $\gamma=1$.
One can see that the concurrence 
starts oscillations when the external 
magnetic field is applied and reaches a limit  
when $~t\to\infty~$. As for the case $\gamma=1$, this limit is not equivalent to 
the concurrence for 
the equilibrium state with $a=b=5.0$. In the lower panel we calculate the thermal
nearest neighbor concurrence $C(i,i+1)$ as a function of time $t$ for $kT=0.5$. \\

Up to now we examined the dynamics of  nearest
neighbor concurrence $C(i,i+1)$ as a function of 
the magnetic filed parameters $(a,b)$, the temperature and 
anisotropy parameter $\gamma$.   To describe the dynamics of the next nearest
neighbor concurrence $C(i,i+2)$, first we compare in Fig. (8) the behavior of
$C(i,i+1)$ and $C(i,i+2)$ as a function of time for same parameters $a=b=1.15$ 
at $kT=0$ and $\gamma=1$. Although $C(i,i+2)$ shows oscillatory behavior as 
for $C(i,i+1)$, but the magnitude is very small compared with the $C(i,i+1)$.
Moreover, by increasing the temperature $C(i,i+2)$ decreases and vanishes for 
$kT > 0.125$   as shown in Fig. (9). In Fig. (10) we show the dynamics of 
$C(i,i+2)$ for $\gamma=0.5$ at $kT=0$ and $kT=0.1$. For this case the value of 
$C(i,i+2)$ is larger than the case with $\gamma=1$ but with similar dynamics.
It is interesting to mention that $C(i,i+2)$ is different
from zero along the magnetic field parameters $a=b$ as shown in Fig. (11)
for $\gamma=1$. The maximum $C(i,i+2) \sim 0.004$ occurs at $(a=1.0,b=1.0)$.
For $\gamma=1$ we need to consider only the dynamics of the nearest neighbor $C(i,i+1)$ and 
next nearest neighbor $C(i,i+2)$ since $C(i,i+3)$ vanishes.

In summary, we have studied the dynamics of entanglement for  one-dimensional spin systems
in an  external magnetic field of a step function form.
We observed that the entanglement shows  nonergodic 
behavior. Due to the coherence of the pairwise 
entanglement with the environment, the change of external magnetic 
field decreases  the nearest and second nearest pairwise entanglement.
 However, at low temperatures, we have found that there are some regions where 
 there is a decoherence of the entanglement due to the change of the 
 external magnetic field. Finally, we have found that  
an increase of the temperature in the 
system will always decrease the pairwise entanglement.

\begin{acknowledgments}

We would like to acknowledge the financial support of the Purdue Research Foundation and
a partial support from the National Science Foundation.

\end{acknowledgments}

\newpage

\begin{figure}
\begin{center}
\includegraphics[width=0.8\textwidth,height=0.8\textheight]{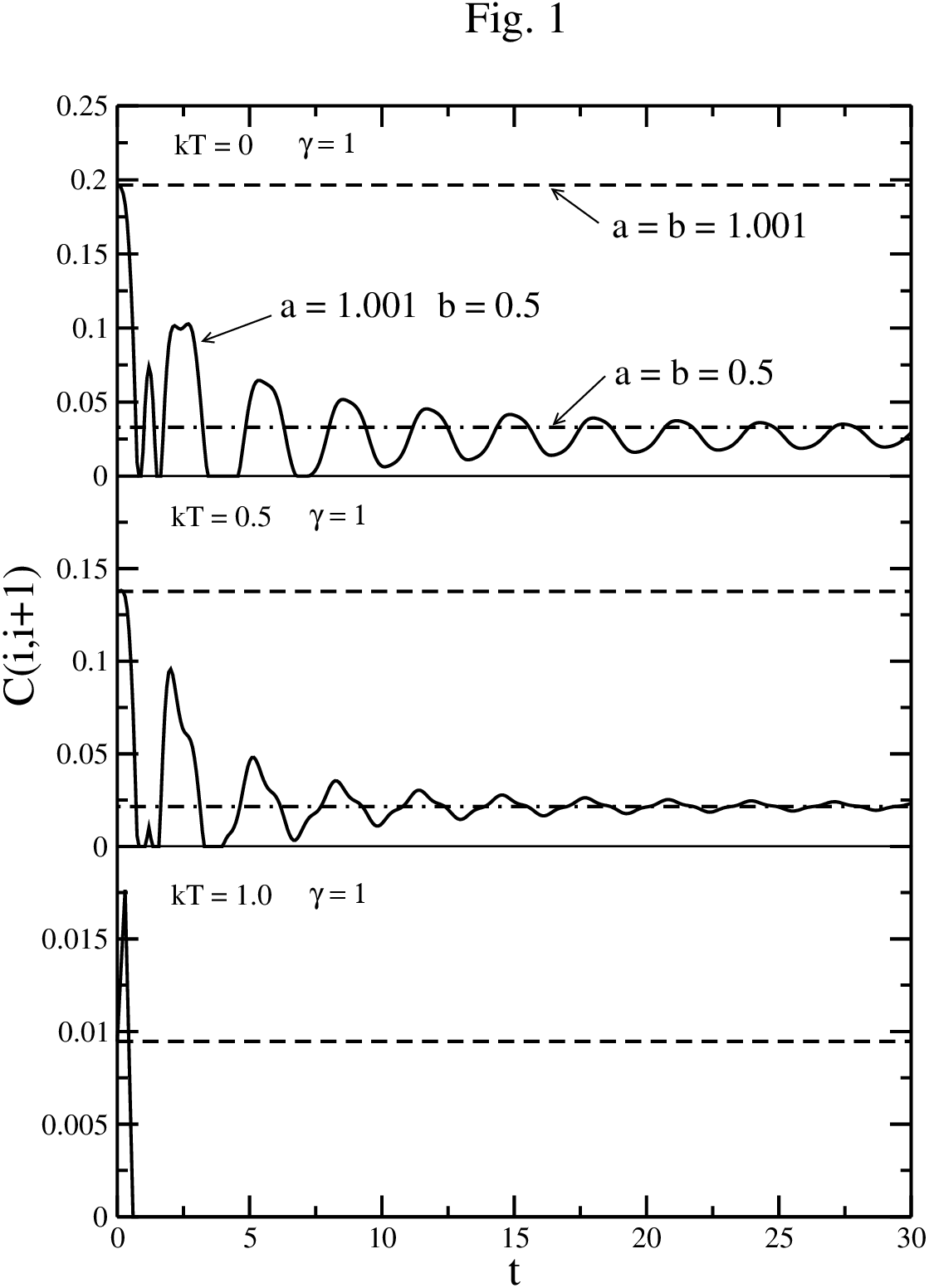}
\end{center}
\caption{The nearest neighbor concurrence $C(i,i+1)$ for different external 
magnetic field strengths $a$ and $b$ as a function of time $t$ for $kT=0$
and $kT=0.5$ and $\gamma=1$.}
 \end{figure}

\newpage

\begin{figure}
\begin{center}
\includegraphics[width=0.8\textwidth,height=0.8\textheight]{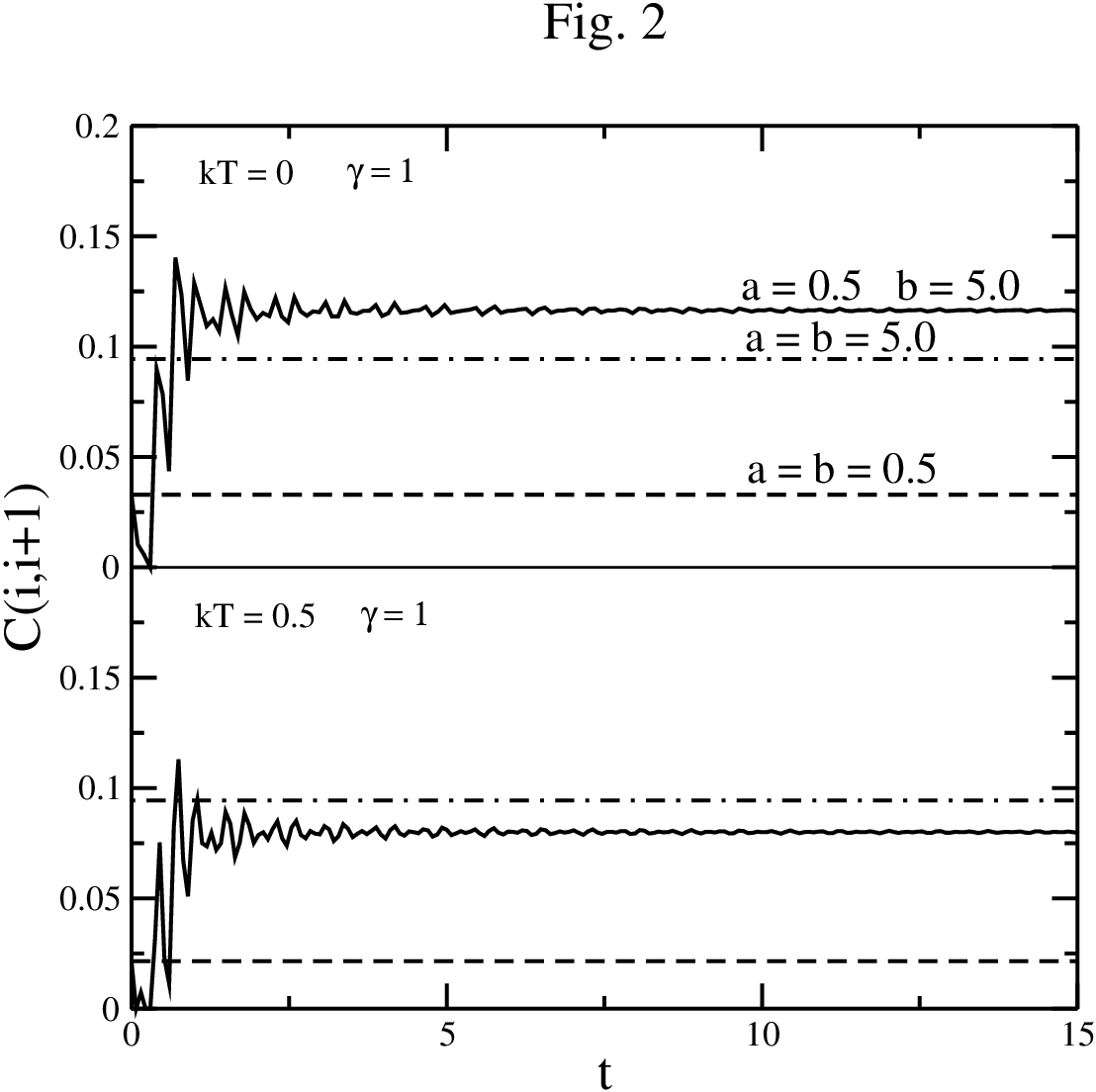}
\end{center}
\caption{The nearest neighbor concurrence $C(i,i+1)$ for different external 
magnetic field strengths $a$ and $b$ as a function of time $t$ for $kT=0$
and $kT=0.5$ and $\gamma=1$.}
 \end{figure}
\newpage

\begin{figure}
\begin{center}
\includegraphics[width=1.0\textwidth,height=0.8\textheight]{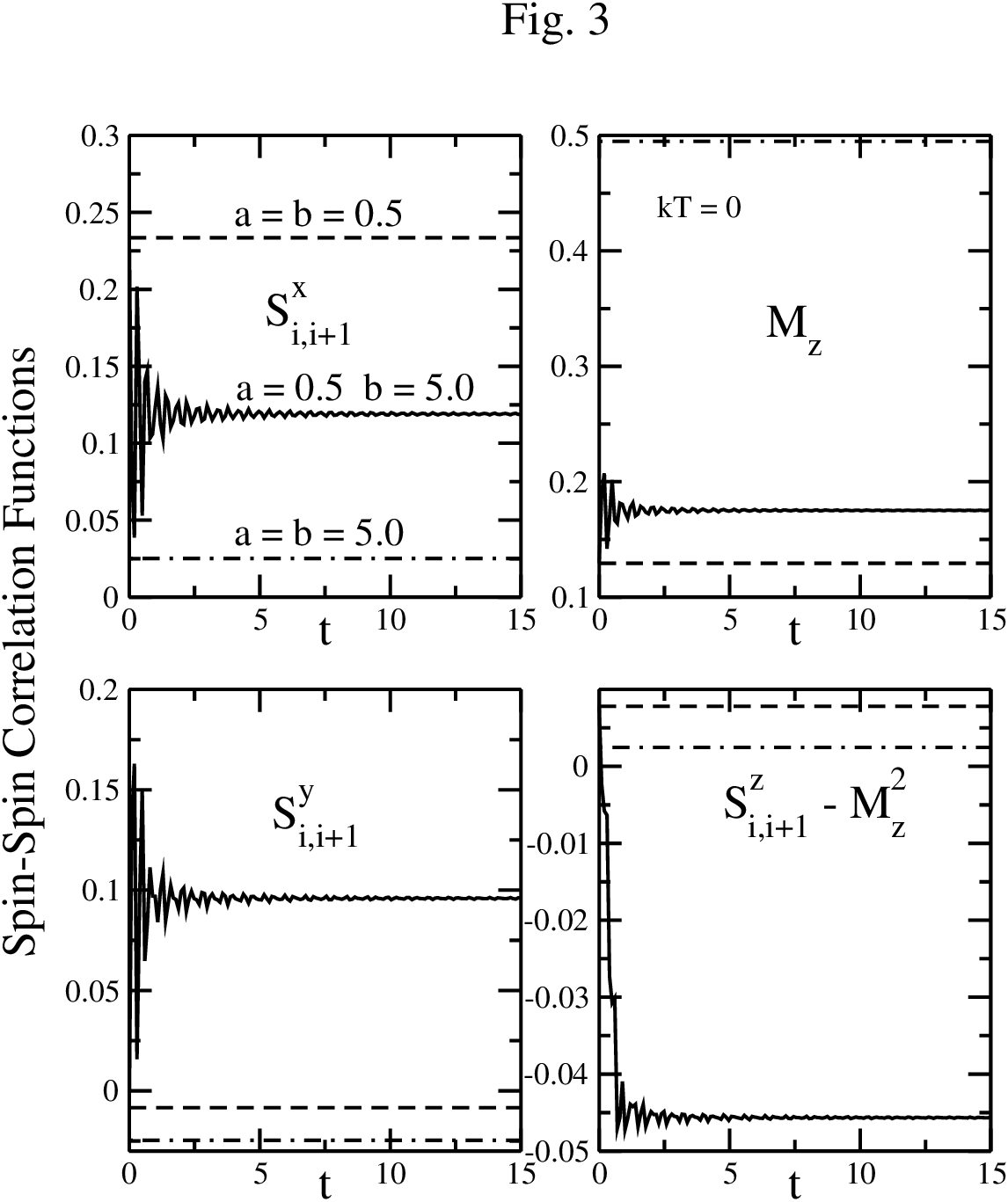}
\end{center}
\caption{The spin-spin correlation functions and the average magnetization 
per spin for different external magnetic field  strengths $a$ and $b$ as a 
function of time $t$ at $kT=0$ for $\gamma=1$.}
 \end{figure}

\newpage

\begin{figure}
\begin{center}
\includegraphics[width=1.0\textwidth,height=0.8\textheight]{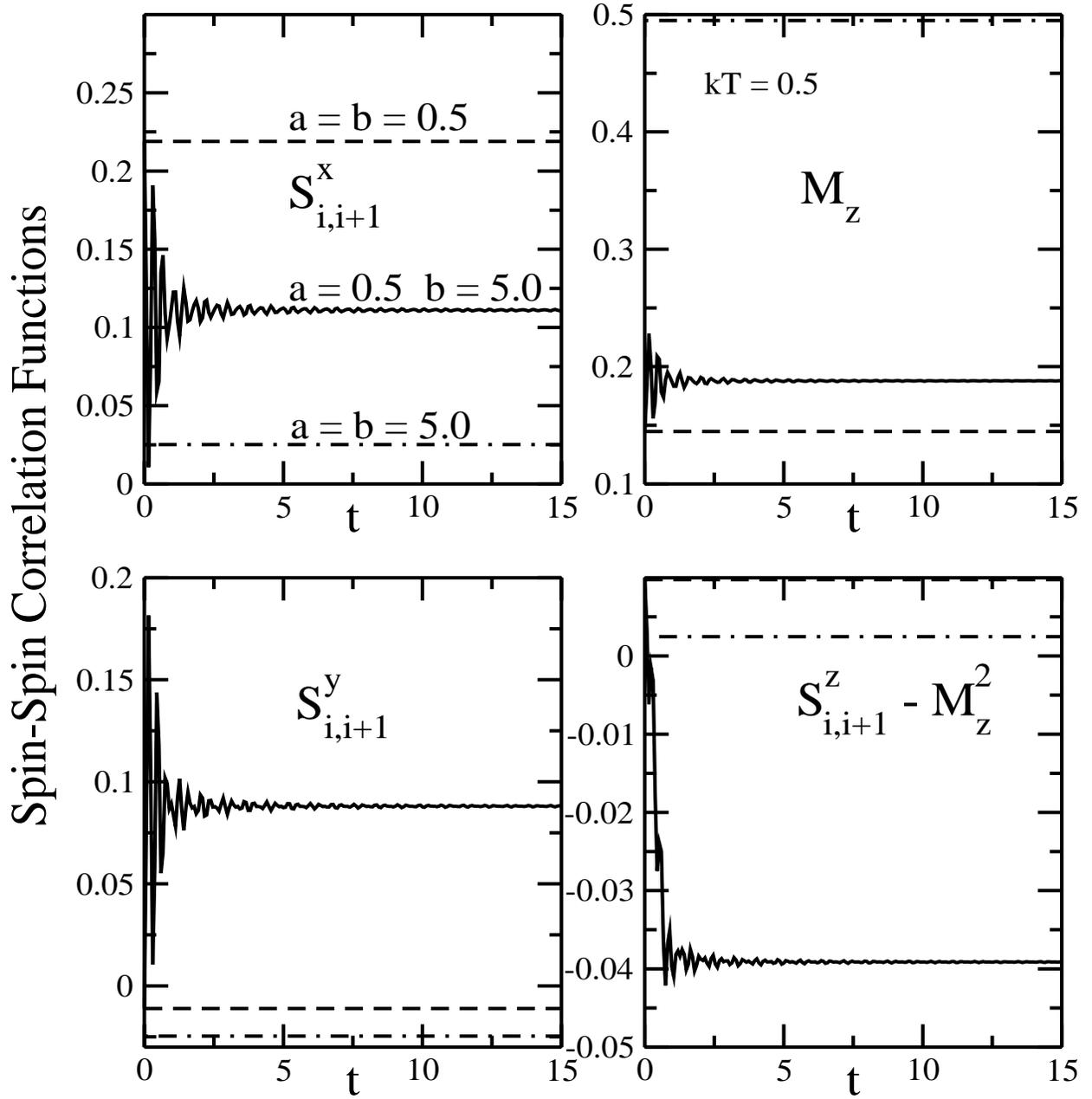}
\end{center}
\caption{The spin-spin correlation functions and the average magnetization 
per spin for different external magnetic field  strengths $a$ and $b$ as a 
function of time $t$ at $kT=0.5$ for $\gamma=1$.}
 \end{figure}

\newpage
\begin{figure}
\begin{center}
\includegraphics[width=1.0\textwidth,height=0.8\textheight]{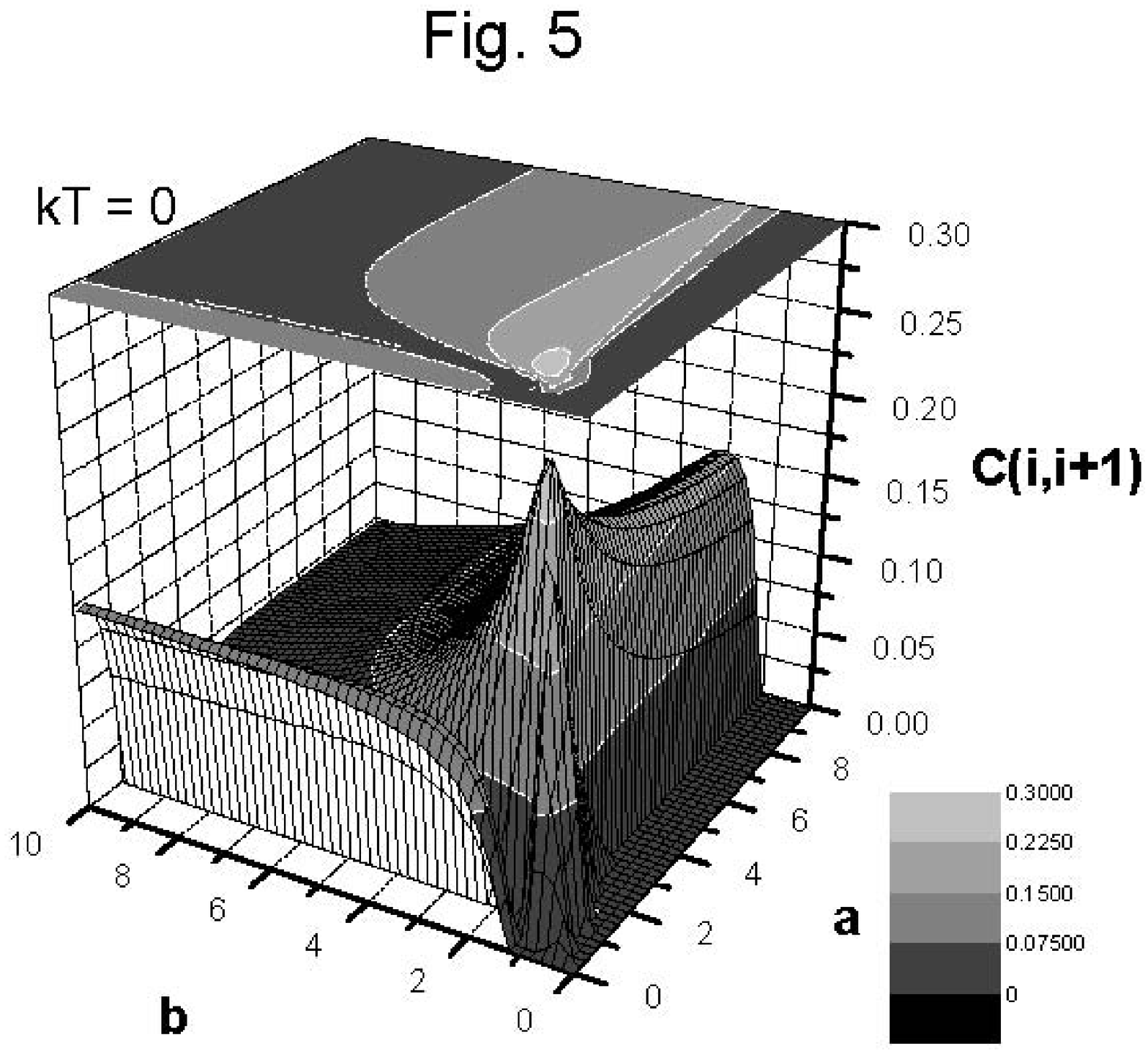}
\end{center}
\caption{The nearest neighbor concurrence $C(i,i+1)$ 
 as functions of different external magnetic field  strengths 
 $a$ and $b$ at time $t\to\infty$ and temperature $kT=0$ for $\gamma=1$.}
\end{figure}

\newpage

\begin{figure}
\begin{center}
\includegraphics[width=1.0\textwidth,height=0.8\textheight]{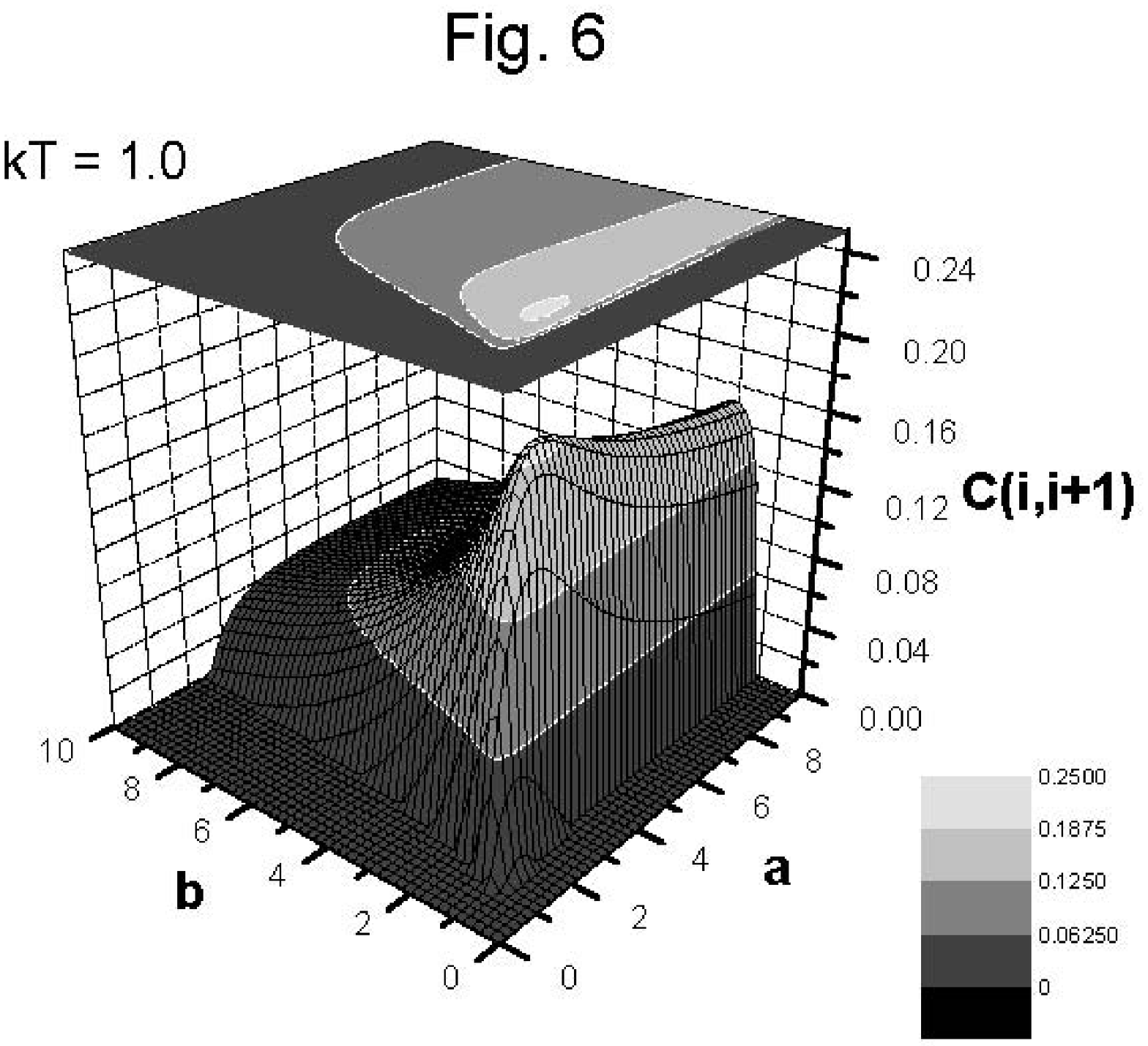}
\end{center}
\caption{The nearest neighbor concurrence $C(i,i+1)$  as functions 
of different external magnetic field  strengths $a$ and $b$ at time 
$t\to\infty$ and temperature $kT=1.0$ for $\gamma=1$.}
 \end{figure}

\newpage

\begin{figure}
\begin{center}
\includegraphics[width=0.8\textwidth,height=0.8\textheight]{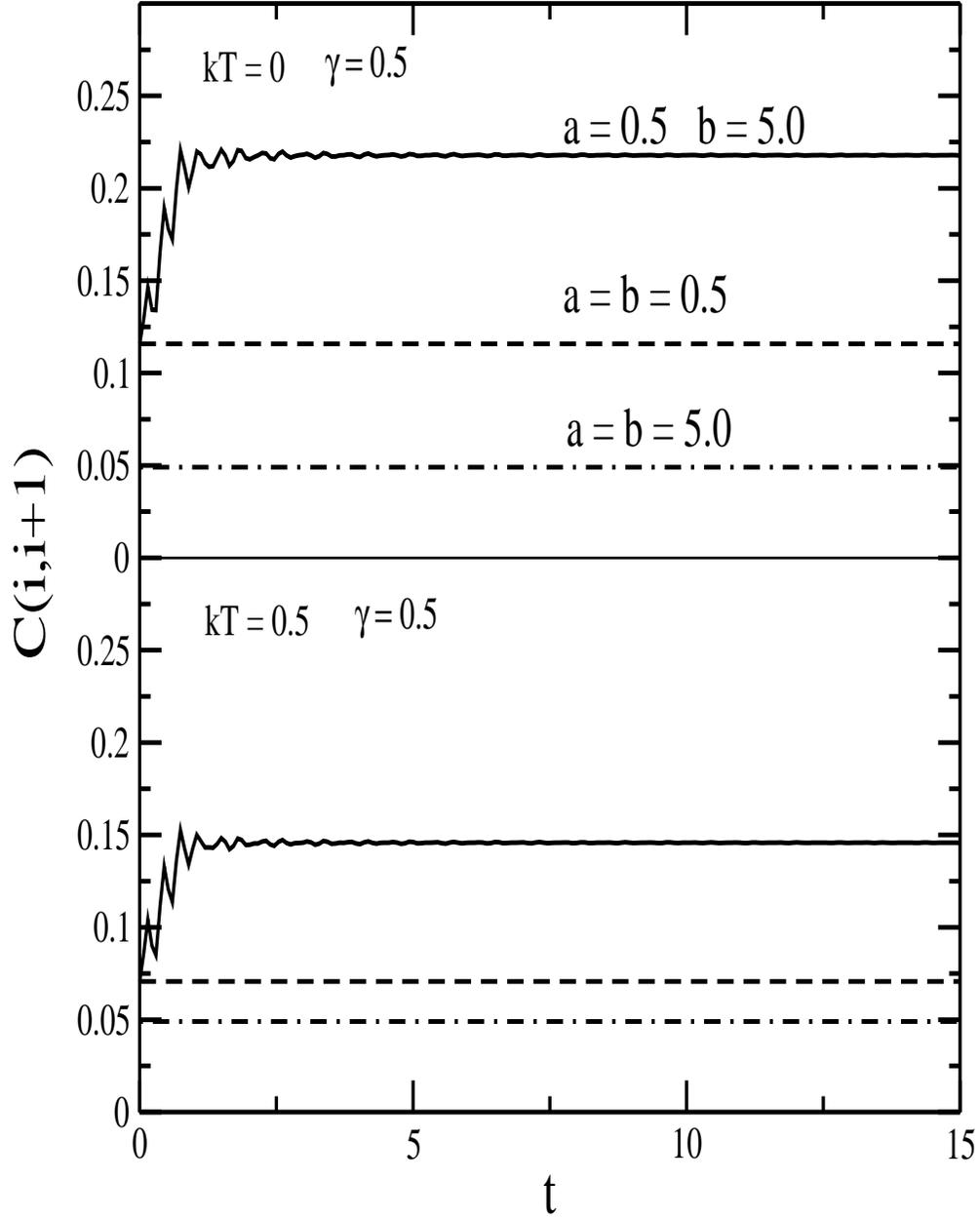}
\end{center}
\caption{The nearest neighbor concurrence $C(i,i+1)$ for different external 
magnetic field strengths $a$ and $b$ as a function of time $t$ for $kT=0$
and $kT=0.5$ and $\gamma=0.5$.}
 \end{figure}

 \newpage

\begin{figure}
\begin{center}
\includegraphics[width=0.8\textwidth,height=0.8\textheight]{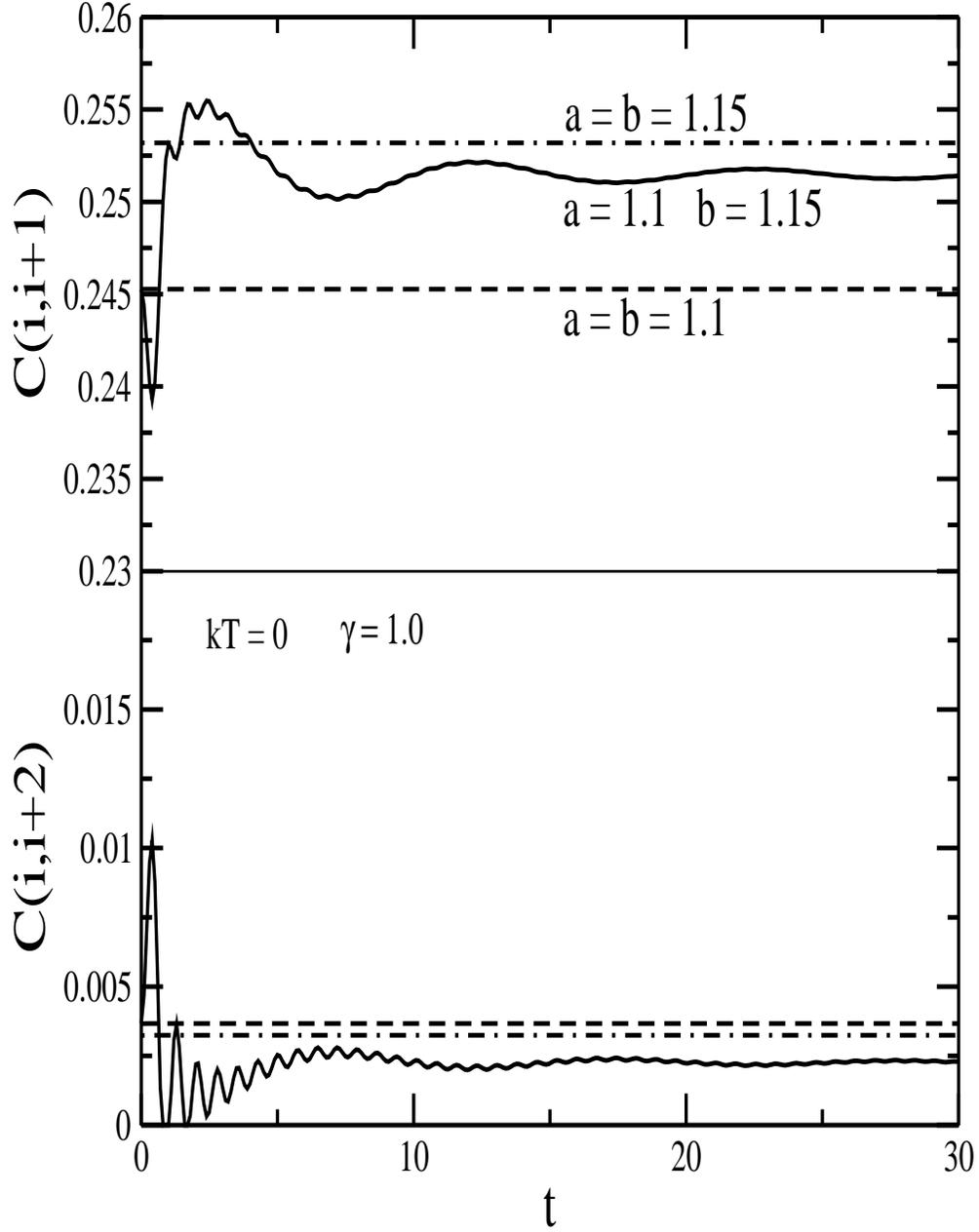}
\end{center}
\caption{Comparison of the nearest neighbor concurrence $C(i,i+1)$ and the 
next nearest neighbor concurrence $C(i,i+2)$ for different external 
magnetic field strengths $a$ and $b$ as a function of time $t$ for $kT=0$
and $\gamma=1$.}
 \end{figure}

\newpage

\begin{figure}
\begin{center}
\includegraphics[width=0.8\textwidth,height=0.8\textheight]{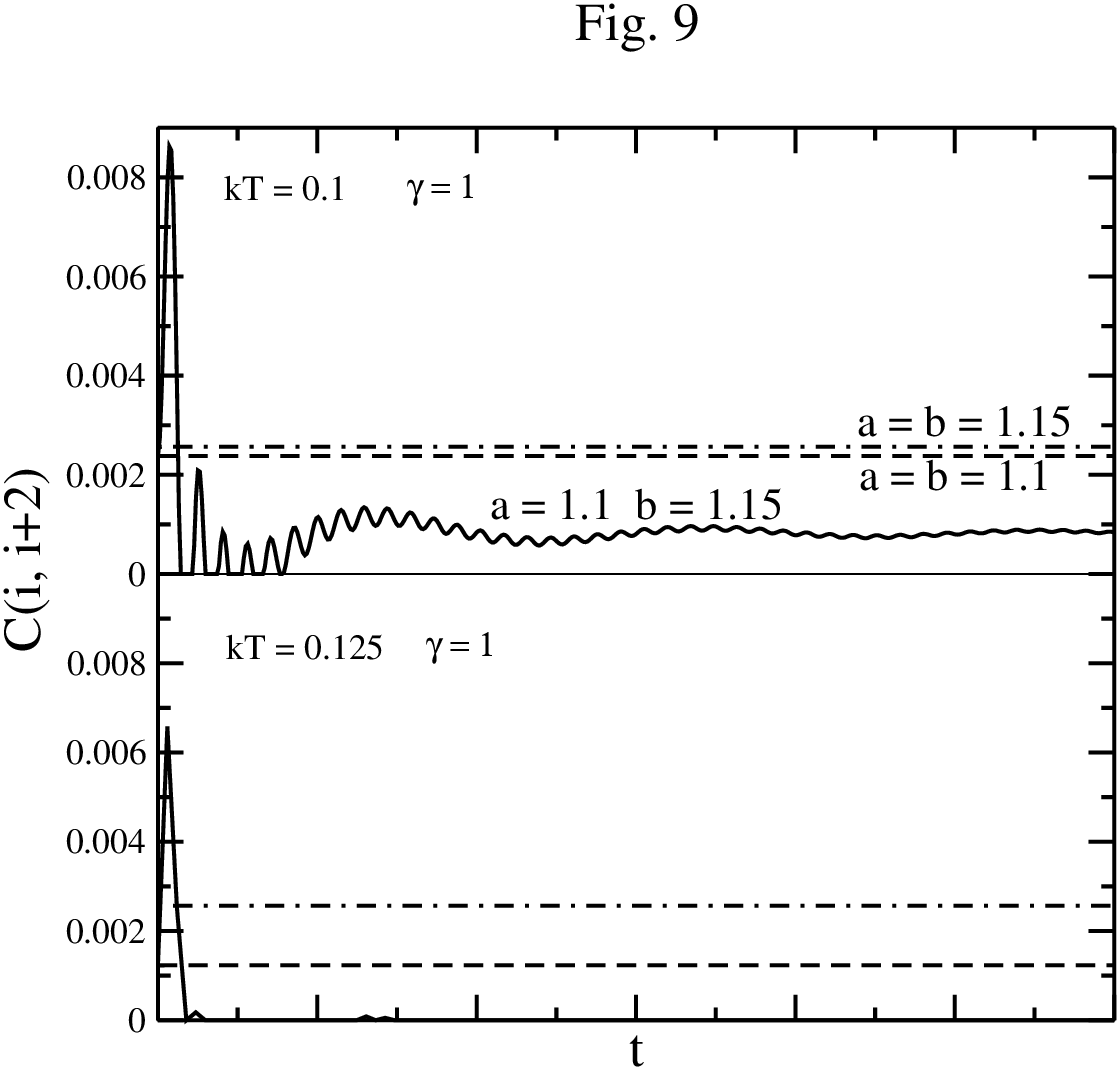}
\end{center}
\caption{The next nearest neighbor concurrence $C(i,i+2)$ for different external 
magnetic field strengths $a$ and $b$ as a function of time $t$ for $kT=0.1$
and $kT=0.125$ for the case $\gamma=1$.}
 \end{figure}

 \newpage

\begin{figure}
\begin{center}
\includegraphics[width=0.8\textwidth,height=0.8\textheight]{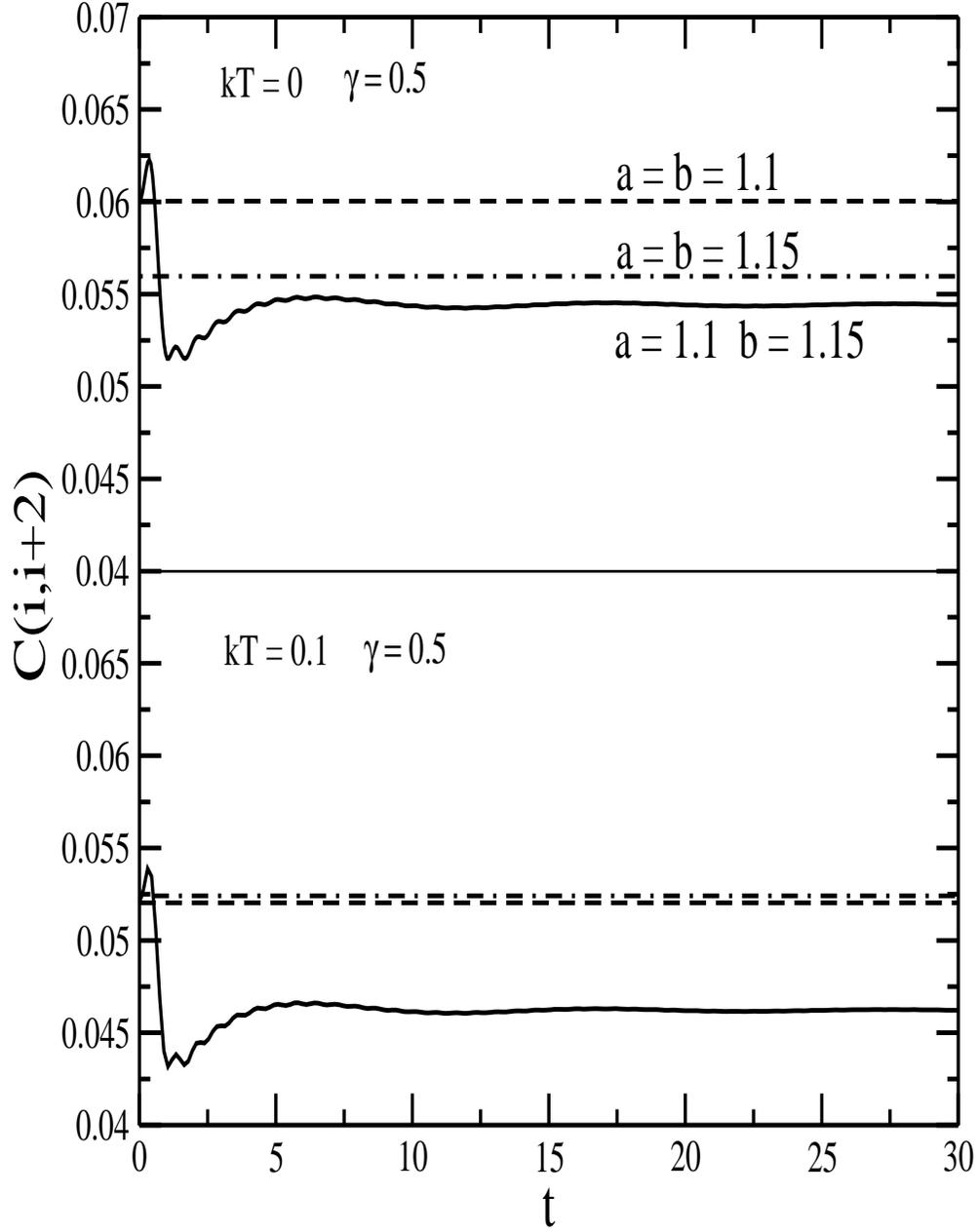}
\end{center}
\caption{The next nearest neighbor concurrence $C(i,i+2)$ for different external 
magnetic field strengths $a$ and $b$ as a function of time $t$ for $kT=0$
and $kT=0.1$ for the case $\gamma=0.5$.}
 \end{figure}
 
\newpage
\begin{figure}
\begin{center}
\includegraphics[width=1.0\textwidth,height=0.8\textheight]{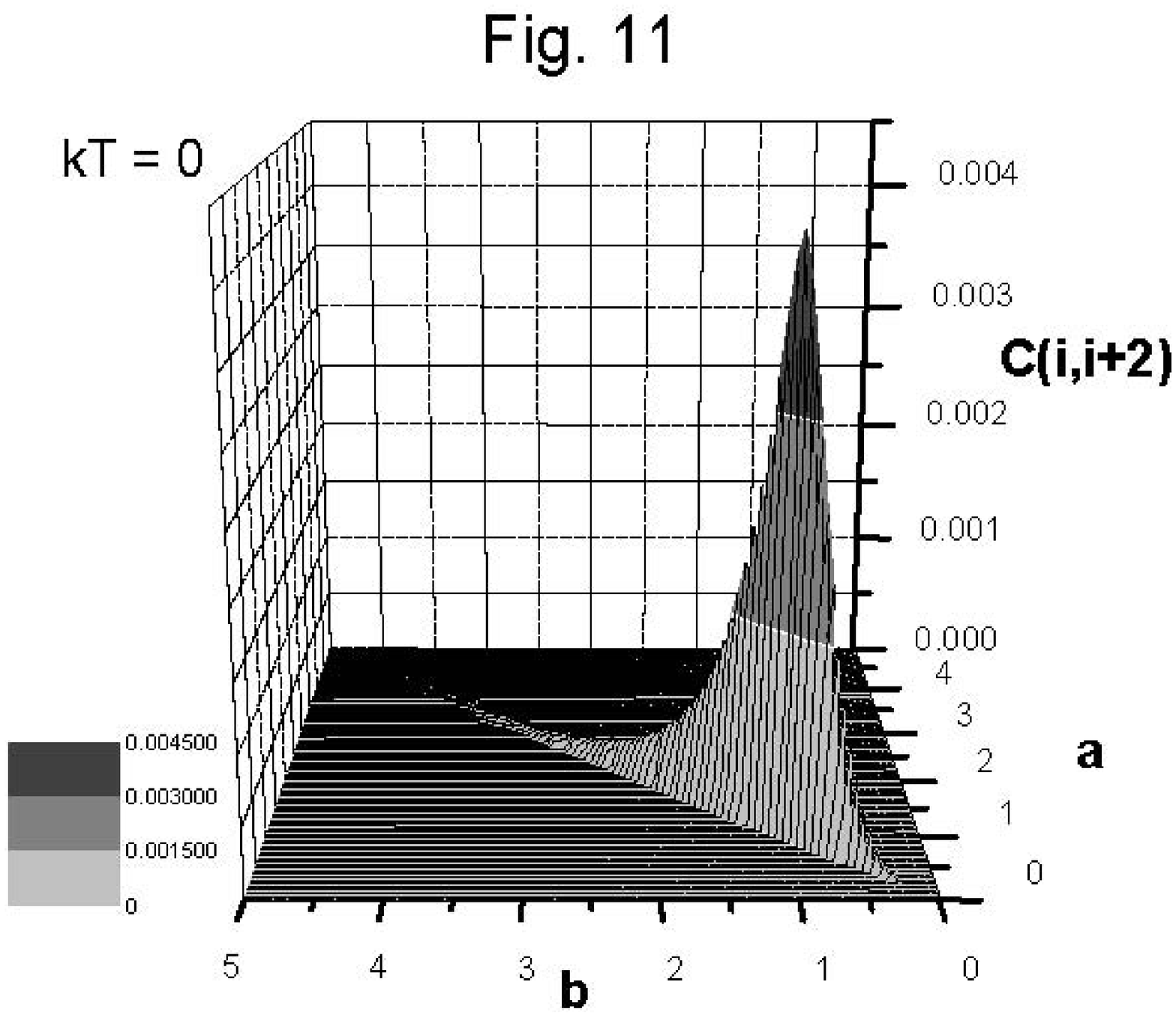}
\end{center}
\caption{The next nearest neighbor concurrence $C(i,i+2)$ 
 as functions of different external magnetic field  strengths 
 $a$ and $b$ at time $t\to\infty$ and temperature $kT=0$ for $\gamma=1$.}
\end{figure}

\newpage

\end{document}